\documentclass[12pt]{article}                              
\usepackage{epsfig}
\usepackage{amsmath,amsthm,amsfonts,amscd,eucal}
\newlength{\dinwidth}
\newlength{\dinmargin}
\newsymbol\rest 1316         

\numberwithin{equation}{section}

\def\cA{{\cal A}}
\def\cB{{\cal B}}

\def\cD{{\cal D}}

\def\cH{{\cal H}}

\def\cO{{\cal O}}

\def\cS{{\cal S}}

\def\bC{{\mathbb C}}

\def\bN{{\mathbb N}}

\def\bR{{\mathbb R}}

\def\g{\gamma}        
\def\d{\delta}        \def\D{\Delta}

\def\l{\lambda}       
\def\m{\mu}
\def\n{\nu}

\def\p{\pi}
\def\r{\rho}
\def\s{\sigma}

\def\t{\tau}

\def\o{\omega}        \def\O{\Omega}

\newtheorem{Thm}{Theorem}[section]
\newtheorem{Cor}[Thm]{Corollary}
\newtheorem{Prop}[Thm]{Proposition}
\newtheorem{Lemma}[Thm]{Lemma}
\theoremstyle{definition}

\theoremstyle{remark}

\begin{document}
\newcommand{\Tmn}{T_{\mu\nu}}
\newcommand{\tmn}{\tau_{\mu\nu}}
\newcommand{\Ain}{{\cal A}_{\infty}}
\newcommand{\Coin}{C^{\infty}_{0}}
\newcommand{\lb}{\mbox{\boldmath $[$}}
\newcommand{\rb}{\mbox{\boldmath $]$}}
\noindent
\begin{center}
{ \Large \bf
        The Averaged Null Energy Condition\\[8pt] for
  General Quantum Field
        Theories\\[14pt] in Two Dimensions}
\\[30pt]
{\large \sc  Rainer Verch}
\\[20pt]
                 Institut f\"ur Theoretische Physik,\\[4pt]
                 Universit\"at G\"ottingen,\\[4pt]
                 Bunsenstr.\ 9,\\[4pt]
                 D-37073 G\"ottingen, Germany\\[4pt]
                 e-mail: verch$@$theorie.physik.uni-goettingen.de
\end{center}
${}$\\[22pt]
${}$ \hfill {\sl Dedicated to the memory of Klaus Baumann}
${}$\\[26pt]
{\small {\bf Abstract. } It is shown that the averaged null energy
  condition is fulfilled for a dense, translationally invariant set of
  vector states in any local quantum field theory in two-dimensional
  Minkowski spacetime whenever the theory has a mass gap and possesses
  an energy-momentum tensor. The latter is assumed to be a Wightman
  field which is local relative to the observables, generates locally
  the translations, is divergence-free, and energetically
  bounded. Thus the averaged null energy condition can be deduced from
  completely generic, standard assumptions for general quantum field
  theory in two-dimensional flat spacetime.}
${}$\\[10pt]
\section{Introduction}
\setcounter{equation}{0}
The averaged null energy condition (ANEC, for short) has attracted
some interest during the past several years as a possible candidate
for a stability condition in semiclassical gravity. In its simplest
form, this condition requires that in quantum field theory (on any
spacetime manifold) the integral of the expectation value, $\langle
\Tmn \rangle$, of the energy-momentum tensor in any physical state,
along any complete, lightlike geodesic $\g$ is always non-negative:
$$ \int_{-\infty}^{\infty} \langle \Tmn(\g(s))\rangle k^{\m}k^{\n}\,ds
\ge 0\,,$$
where $s$ is an affine parameter and $k^{\m}$ the (parallelly
propagated) tangent of $\g$. (For a formulation not
requiring the existence of the integral, see below.)

We shall briefly indicate the origin and development of this
condition, however, we are
not attempting to properly review this area of research and refer the
reader to the articles \cite{FlaWa,Y1,Y2,FR,WalYu} for further discussion
and additional references.

In the theory of classical gravity, one central object of study is the
behaviour of solutions to Einstein's equations,
$$ G_{\m \n}(x) = 8\p \Tmn(x)\,,$$
for classical matter described by the energy-momentum tensor
$\Tmn$. There are important results asserting that a certain
qualitative behaviour of these solutions must necessarily occur,
within a broad range of initial conditions, as soon as certain
stability requirements are imposed on $\Tmn$. It is significant that
such qualitative behaviour typically reflects a stability of
causality, i.e.\ an initially causally well-behaved spacetime will not
end up to develop, e.g., closed timelike curves. Most prominent among
those results are the singularity theorems \cite{HE,WaldI}; the
typical stability requirements in this context are the null energy
condition,
\begin{equation}
 \Tmn(x)k^{\m}k^{\n} \ge 0
\end{equation}
for all lightlike vectors $k^{\m}$ at any point $x$ in spacetime, or
the weak energy condition, where (1.1) is to hold for all causal
vectors $k^{\m}$ at any point $x$, and related variants, like the
strong energy condition or the dominant energy condition, cf.\
\cite{HE,WaldI}.
The common feature of these conditions is that they impose a local
(even pointlike) positivity constraint like in eq.\ (1.1) on the
energy-momentum tensor. For energy-momentum tensors of
phenomenological models for classical matter, such local positivity
constraints have largely been found to be physically realistic. In
contrast, it is known that, under very general hypotheses, similar
local positivity constraints cannot hold for the expectation values of
the energy-momentum tensor $\langle
\psi,\Tmn(x)\psi \rangle$ of a quantum field in Minkowski-spacetime
for a dense set of state vectors $\psi$ \cite{EGJ}.

Now, in semiclassical gravity, one investigates the semiclassical
Einstein equation
\begin{equation}
 G_{\m \n}(x) = 8\p\langle \Tmn(x)\rangle
\end{equation}
where $\langle \Tmn(x)\rangle$ is the expectation value of the
energy-momentum tensor in a physical state of a quantum field
propagating in a classical background spacetime whose Einstein tensor
is $G_{\m \n}$. The question arises if there is a realistic replacement
for the local positivity constraints on $\langle \Tmn(x)\rangle$
leading to similar implications, i.e.\ the necessity of a certain,
causally stable behaviour of solutions to (1.2) to occur. And in fact,
candidates for such replacements have been found. In \cite{Tip} it was
observed that nonlocal, ``averaged'' versions of the local positivity
constraints on the classical energy-momentum tensor still lead to
essentially the same singularity theorems which result from imposing
local positivity constraints (see also \cite{ChiEh,Borde,Rom} for
discussion and further results). The ``averaged'' refers to
integrating the energy-momentum tensor along causal geodesics. The
condition used in \cite{Tip} is that
$$ \int_{-\infty}^{\infty} \left(\Tmn -
 \mbox{$ \frac{1}{2}$} g_{\m \n}T^{\s}{}_{\s}\right)\!\!(\g (s))
\, k^{\m}k^{\n}\,ds \ge 0 $$
for any complete causal geodesic with affine parameter $s$ and tangent
$k^{\m}$; $g_{\m \n}$ is the spacetime metric. This is referred to as
averaged strong energy condition. In \cite{Rom} it was shown that an
averaged null energy condition for certain half-complete geodesics,
i.e.\ essentially
\begin{equation}
 \liminf_{r \to \infty}\, \int_0^{r}\Tmn(\g (s))\,k^{\m}k^{\n}\,ds \ge
 0
\end{equation}
for all lightlike geodesics $\g$ with affine parameter $s$ and tangent
$k^{\m}$ emanating at $s =0$ from a closed trapped surface, implies
singularity theorems. Moreover, it is proved in \cite{Borde} that
singularity theorems are implied by ANEC, roughly,
\begin{equation}
\liminf_{r_{\pm} \to \infty}\,\int_{-r_-}^{r_+} \Tmn(\g
(s))\,k^{\m}k^{\n}\,ds \ge 0
\end{equation}
for all complete lightlike geodesics with affine parameter $s$ and
tangent $k^{\m}$. (The precise formulations in \cite{Rom} and
\cite{Borde} are slightly different from ours in (1.3) and (1.4). The
reader is referred to these references for the technical details. The
significant point is that the averaged energy conditions don't assume that the
integrals converge, nor that they are bounded above.)
It was also shown in \cite{MTY} and \cite{FSW} that the averaged null
energy conditions (1.3) and (1.4), respectively, prevent the occurence
of traversable wormholes in solutions to Einstein's equations.

In the light of these findings, an interesting issue is whether such
averaged energy (positivity) conditions are fulfilled for the
expectation values of the energy-momentum tensor for (suitable) states
in quantum field theories. There have been several works dealing with
this question and we continue by summarizing, however briefly, the
results found so far.
To fix our terminology, we say that a state $\o$ of a quantum field
theory on some background spacetime fulfills the ANEC (resp., AWEC =
averaged weak energy condition) if the expectation value $\langle
\Tmn(x)\rangle_{\o}$ of the energy-momentum tensor
exists in this state as a (smooth) function of
all $x$ in spacetime, and if
\begin{equation}
 \liminf_{r_{\pm} \to \infty}\, \int^{r_+}_{-r_-}\langle \Tmn(\g (s))
 \rangle_{\o}\, k^{\m}k^{\n}\, ds \ge 0
\end{equation}
 holds for all complete lightlike (resp., timelike) geodesics $\g$
 with affine parameter $s$ and tangent $k^{\m}$. (However, it should
 be noted that in some references slightly different formulations are
 used.)

In \cite{Klin} it is shown that ANEC and AWEC are fulfilled for the
free scalar field in $n$-dimensional Minkowski spacetime for states
which are bounded in particle number and energy. It was also found in
this work that AWEC is violated in some states of the free scalar field
on a static, spatially closed two-dimensional spacetime.

The work \cite{Fol} establishes ANEC for states bounded in particle
number and energy of the free electromagnetic field in
four-dimensional Minkowski spacetime.

In the article \cite{WalYu} it is shown that ANEC holds for all
Hadamard states of the massless free scalar field in any
two-dimensional globally hyperbolic spacetime, and for all Hadamard
states of the massive free scalar field on two-dimensional Minkowski
spacetime. Moreover it is proved that ANEC holds for the massive and
massless free scalar fields fulfilling some additional condition
(implying that the limit $r_{\pm} \to \infty$ in (1.5) exists)
in four-dimensional flat spacetime. In
that work there appears also an argument indicating that ANEC cannot
be expected to hold in general for the massless free scalar field on
all four-dimensional curved spacetimes. Conditions implying that ANEC
and AWEC will fail to hold generally for a large class of curved
spacetimes are given in \cite{Vis}. In \cite{Y1,Y2} it has therefore
been suggested that the original formulation of ANEC should be altered
via replacing the integrand of (1.5) by 
$$ \langle \Tmn(\g (s))\rangle_{\o} - D_{\m \n}(\g (s)) $$
where $D_{\m \n}(x)$ is some state-independent tensor, e.g.\ the
expectation value of the energy-momentum tensor in some reference
state (like the vacuum in flat spacetime) or some quantity locally
constructed from curvature terms. Such formulation of ANEC has been
termed ``difference inequality''. Results in \cite{Y1,Y2} and
\cite{FR} (cf.\ also \cite{FlaWa}) indicate that such difference
 inequalities may have a better
chance to hold generally in curved spacetime. We refer to the
references for further discussion.

At any rate, investigations about the validity of ANEC (or difference
inequalities) so far have been limited to the consideration of free
fields only. The  proofs of ANEC presented up to now
strongly rely either on the fact that the quantum field obeys a linear
hyperbolic equation of motion, or on the explicit form of the Wick-ordered
energy-momentum tensor operator as bilinear expression in
annihilation and creation operators in Fockspace. This is clearly
unsatisfactory if one wishes to assess the general validity of ANEC in
quantum field theory (say, in flat spacetime). Moreover, one would
like to understand the connection of ANEC to the standard stability
requirement in general quantum field theory, i.e.\ the spectrum
condition and existence of a vacuum state.

In the present work, we make a first attempt towards clarifying the
status of ANEC in general quantum field theory. We shall consider a
general quantum field theory on two-dimensional Minkowski spacetime
obeying the usual assumptions like locality, translation covariance,
spectrum condition with mass gap and existence of a unique
vacuum. Furthermore we assume that such a theory possesses an
energy-momentum tensor, which is essentially supposed to be a Wightman
field (operator valued distribution) characterized by being local
relative to the observables, divergence-free, generating locally the
translations, and fulfilling an energy-bound. The precise assumptions
are given in Section 2.1. Comments on these assumptions and some
well-known consequences (needed later) appear in Section 2.2. In
Section 2.3 we prove that ANEC is fulfilled for a dense,
translationally invariant set of vector states of any quantum field
theory in two-dimensional Minkowski-spacetime fulfilling the general
assumptions of Section 2.1. In Section 3 we show that ANEC will in
general fail to hold if the integral averaging is carried out only
along a lightlike geodesic half-line as in (1.3). This is of course
expected in view of locality and the Reeh-Schlieder property. Some
concluding remarks appear in Section 4.

We have opted to stage our discussion in the framework of the
operator-algebraic approach to local quantum field theory (cf.\
\cite{Haag,BW}) since this makes the structures involved in the
argument particularly transparent. One could also obtain similar
results working entirely in the setting of Wightman fields \cite{SW}.

\section{ANEC in quantum field theory on two-dimensional Minkowski
  spacetime}
\setcounter{equation}{0}
\subsection{Assumptions}
Our discussion of the ANEC in general quantum field theory on
two-dimensional Minkowski spacetime begins by formulating the relevant
assumptions.
\\[6pt]
{\it Notation. } Two-dimensional Minkowski-spacetime will be
identified, as usual, with $\bR^2$ with metric $(\eta_{\m\n}) = {\rm
  diag}(1,-1)$. The open forward lightcone is the set $V_+ :=\{x \in
\bR^2: (x^0)^2 - (x^1)^2 > 0,\ x^0 > 0\}$, the open backward lightcone
is $V_- := - V_+$. The causal complement, $\cO^{\perp}$,
 of a set $\cO \subset \bR^2$ is the largest open complement of the
 union of all sets $(V_+ \cup V_- )
  + x$, $x \in \cO$. A {\it double cone} is a set of the form
$\cO_{I} := (S\backslash I)^{\perp}$ where $S$ is any spacelike line
in $\bR^2$ (a spacelike hypersurface) and $I$ any finite open
subinterval of $S$. Any double cone is of the form $\cO = (V_+ + y)
\cap (V_- + x)$ for pairs of points $x,y \in \bR^2$ with $x \in V_+
+y$. A {\it wedge region} is of the form $W = L(W_R)$ for any
Poincar\'e transformation $L$ where $W_R$ is the right wedge, $W_R
:=\{(x^0,x^2) \in \bR^2 : 0<x^1,\ |x^0| < x^1 \}$.

There will often appear the following special elements in $\bR^2$:
$$
e_0 := \left(^1_0\right)\,,\ \ e_1 := \left(^0_1\right)\,, \ \ e_+ :=
\mbox{$\frac{1}{\sqrt{2}}$} (e_0 +
e_1)\,,\ \  e_- := \mbox{$\frac{1}{\sqrt{2}}$}(e_0 - e_1)\,. $$

 The summation convention is used throughout.
\\[6pt]
We shall now define what we mean by a quantum field theory with an
energy-momentum tensor on two-dimensional Minkowski-spacetime: This is
described in terms of a collection of objects $\{\cH,\cA,U,\O,\Tmn\}$
whose properties are assumed to be as follows:
\begin{itemize}
\item[(i)] $\cH$ is a Hilbertspace, and there is a map $\cO \mapsto
  \cA(\cO)$ assigning to each double cone $\cO$ in $\bR^2$ a von
  Neumann algebra in $\cB(\cH)$, with the properties:
\\[2pt]
${}$ \quad $\tilde{\cO} \subset \cO \quad \  \Rightarrow \quad
\cA(\tilde{\cO}) \subset 
\cA(\cO)$ \quad (isotony),
\\[2pt]
${}$ \quad $\tilde{\cO} \subset \cO^{\perp} \quad \Rightarrow \quad
\cA(\tilde{\cO}) \subset\cA(\cO)'$ \quad (locality).
 \footnote{Recall that $\cA(\cO)'$ is
  the commutant of $\cA(\cO)$, i.e.\ the algebra formed by all
  operators in $\cB(\cH)$ that commute with every element in
  $\cA(\cO)$.}
\item[(ii)] There is a weakly continuous representation $\bR^2 \owns a
  \mapsto U(a)$ of the two-dimensional translation group by unitary
  operators on $\cH$, fulfilling for all double cones $\cO$,
$$ U(a)\cA(\cO)U(a)^* = \cA(\cO + a)\,, \quad a \in \bR^2
\quad {\rm (covariance)}.$$
\item[(iii)] There is an up to a phase unique unit vector $\O \in
  \cH$ which is left invariant by the unitary group $U(a)$, $a \in
  \bR^2$ \quad (existence of a unique vacuum).
\item[(iv)] Denote by $P = (P_0,P_1)$ the generator of $U(a)$, $a \in
  \bR^2$, i.e.\ $U(a) = {\rm e}^{iP_{\m}a^{\m}}$. Its spectrum
  fulfills
\\[2pt]
$${\rm sp}(P) \subset \{0\} \cup \{(p_0,p_1) \in \bR^2:
(p_0)^2 -(p_1)^2 \ge m > 0, \ p_0 > 0\} $$
for some fixed $m > 0$ \ \,(spectrum condition with mass gap).
\item[(v)] The vacuum vector $\O$ is cyclic for union of the local von Neumann
  algebras $\bigcup_{\cO} \cA(\cO)$,
 i.e.\  the set $\bigcup_{\cO}\cA(\cO)\O$ is
  dense in $\cH$ \quad (cyclicity of the vacuum).
\item[(vi)] We denote by $\Ain$ the $*$-subalgebra of $\cB(\cH)$
  generated by all operators $A$ of the form
$$A = \int h(a)\,U(a)B U(a)^*\, d^2a$$
for $h \in \Coin(\bR^2)$ and $B \in \bigcup_{\cO} \cA(\cO)$, and
define: 
$$\Ain(\cO) := \Ain \cap \cA(\cO)\,.$$
\par
The energy-momentum tensor, $\Tmn$, $\n , \m = 1,2$, is a set of
operator valued distributions; more precisely, there is a dense domain
$D \subset \cH$, with $U(a)D \subset D$, $a \in \bR^2$, and $\Ain\O
\subset D$, so that for each $f \in \Coin(\bR^2)$, $\Tmn(f)$ is a
closable operator on $D$ with $\Tmn(\overline{f}) \subset
\Tmn(f)^*$. For each $\psi,\psi' \in D$, the map 
$$\Coin(\bR^2) \owns f \mapsto \langle \psi,\Tmn(f) \psi'
\rangle$$
is a distribution in $\cD'(\bR^2)$.
\item[(vii)] Translation-covariance holds:
$$U(a)\Tmn(f)U(a)^* = \Tmn(f_a)\,, \quad f \in
\Coin(\bR^2),\ a \in \bR^2\,,$$
with $f_a(x) := f(x -a)$. Moreover, $\Tmn$ has vanishing
vacuum-expectation value:
$$\langle \O,\Tmn(f)\O \rangle = 0\,, \quad f \in
\Coin(\bR^2)\,. $$
\item[(viii)] $\Tmn$ is local on the vacuum:
\quad $\langle A\O,\lb \Tmn(f),B \rb \O \rangle = 0$ \\[2pt]
for all $A \in \Ain$, $B \in \Ain(\cO)$ and $f \in \Coin(\cO^{\perp})$.  
\item[(ix)] $\Tmn$ is divergence-free on the vacuum:
$$ \langle A \O,\lb \Tmn(\partial^{\m}f),B \rb \O \rangle = 0\,, \quad
A,B \in \Ain \,.$$
\item[(x)] $\Tmn$ generates (locally) the translations on the vacuum:
  Let $S$ be the
  $x^0 = 0$ hyperplane (= spacelike line) with unit normal vector
  $e_0$.
Whenever $a,b \in \Coin(\bR)$ are any two non-negative functions with
the properties
\\[2pt]
${}$ \quad $a(x^0) =0$ outside of some $x^0$-interval
$(-\varepsilon_a,\varepsilon_a)$ and $\int a(x^0)\,dx^0 = 1$,
\\[2pt] 
${}$ \quad $b(x^1) = 1$ on an open $x^1$-interval $(\xi_b
-\varepsilon_a -\d_b,\xi_b + \varepsilon_a + \d_b)$,
\\[2pt]
${}$ \quad where $\varepsilon_a,\d_b > 0$, $\xi_b \in \bR$,
\\[2pt]
we require that, upon setting $\chi(x^0,x^1) := a(x^0)b(x^1)$,
there holds
\begin{eqnarray*}
 \langle A \O,\lb \Tmn(\chi),B\rb \O \rangle e_0^{\m} & = & \langle A
 \O,\lb P_{\n},B \rb \O \rangle \\
& = & \langle A \O,P_{\n}B \O \rangle
\end{eqnarray*}
for all $A \in \Ain$ and all $B \in \Ain(\cO_I)$ with $I = (\xi_b
-\d_b,\xi_b + \d_b)$.
\item[(xi)] Energy bounds for $\Tmn$: \quad There is a pair of numbers
  $c,\ell >0$ such that $(1 + P_0)^{-\ell}\Tmn(f)(1+P_0)^{-\ell}$ is for
  each $f \in \Coin(\bR^2)$ a bounded operator whose operator norm
  satisfies the estimate
$$ ||\,(1 + P_0)^{-\ell}\Tmn(f)(1+P_0)^{-\ell}\,|| \le
c\,||\,f\,||_{L^1}\,, \quad f \in \Coin(\bR^2)\,.$$
\end{itemize}
\subsection{Comments and some implications}
The conditions (i)--(v) imply that we are considering a
translation-covariant quantum field theory in a vacuum representation
with mass gap, in operator algebraic formulation. These conditions are
quite standard; the selfadjoint elements in $\cA(\cO)$ are viewed as
observables of the theory localized in the spacetime region $\cO$,
cf.\ \cite{Haag} for further discussion. 

Note that (v)  and uniqueness of the vacuum vector 
imply irreducibility of the observable algebra, i.e.\
$(\bigcup_{\cO}\cA(\cO))' = \bC\,1$. Note also that locality, spectrum
conditon and (v) imply the Reeh-Schlieder property of the algebras
associated with wedge-regions $W$, defined as $\cA(W) := (\bigcup_{\cO
  \subset W}\cA(\cO) )''$, i.e.\ the sets  $\cA(W)\O$ are dense in
$\cH$ for
any wedge-region. It follows easily that then also the sets
$\cA_{\infty}(W)\O$ are dense in $\cH$ for all wedge regions $W$.
A slightly stronger assumption would be the Reeh-Schlieder property
for the local algebras, i.e.\ that $\cA(\cO)\O$ is dense in $\cH$ for
each double cone $\cO$;
this is the case when the local von Neumann algebras are weakly
additive, as e.g.\  when  there is a Wightman field
generating the local algebras \cite{ReS,SW}. Then it follows that
$\Ain(\cO)\O$ is dense in $\cH$. We will make such an assumption in Section
3.

The conditions (vi)--(xi) serve to characterize an energy-momentum
tensor in the present abstract setting. Conditions (vi)--(viii)
basically say that the energy-momentum tensor is a Wightman field
which is local relative to the observables. Particularly important for
the interpretation of $\Tmn$ as an energy-momentum tensor are clearly
(ix) and (x) expressing that, in a weak sense, $\Tmn$ is
divergence-free and generates locally the translations. Notice that on
account of the assumed translation-covariance the condition formulated
in (x) implies its validity for any translated copy $S +a$, $a \in
\bR^2$, of $S$ in place of $S$. It is worth pointing out that we could
have also taken for $S$ any other spacelike hyperplane (= spacelike
line) instead of the $x^0 = 0$ hyperplane, the proof of
Theorem 2.5 below would then only involve changes in notation. Specializing
to the $x^0 = 0$ hyperplane is thus just a matter of notational
convenience.

Notice
that for each $A \in \Ain$ one has $A\O \in \bigcap_{j \in \bN}{\rm
  dom}\,(1+ P_0)^j$.
In view of the assumed energy bound, it actually follows that $\Ain\O$
is contained in the domain of $\Tmn(f)$. (An assumption of this kind
is clearly needed, otherwise it would be difficult to formulate that
$\Tmn(f)$ is local relative to the observables.) The energy bound (xi)
has the simple interpretation that the local energy-momentum density
integrated over a finite spacetime volume should be dominated by the
total energy (or at least a sufficiently high moment of it). We
mention as an aside that, if one assumes the domain $D$ of $\Tmn$ to
coincide with the set $\bigcap_{j \in \bN}{\rm dom}\,(1+P_0)^j$ and
takes as testfunction-space the Schwartz-functions $\cS(\bR^2)$
instead of $\Coin(\bR^2)$, then this implies already an energy bound
of the form as in (xi) \cite[Prop. 12.4.10]{BW}.

Finally, there arises the question if our assumptions regarding $\Tmn$
are realistic. For free fields, the canonically constructed
energy-momentum tensor fulfills the assumptions. (It fulfills, in
particular, a linear energy bound, i.e.\ (xi) holds with $\ell = 1$.)
While we have made no attempt to check this, 
it is to be expected that the quantum field models which have been
constructed in two dimensions, like $P(\phi)_2$ or Yukawa$_2$, also
comply with all of our assumptions.
\\[6pt]
The assumptions (i)--(xi) for a theory with energy-momentum tensor,
$\{\cH,\cA,U,\O,\Tmn\}$, are known to imply certain properties which
will be used in deriving ANEC in the next section. Hence we
subsequently collect these properties, mainly referring to the
literature for proofs.
\begin{Prop} {\rm \cite{Bor,Dri}} One has weak asymptotic lightlike
  clustering: For any lightlike $k \in \bR^2\backslash\{0\}$ and any
  pair of vectors $\psi,\psi' \in \cH$, it holds that
\begin{equation}
\lim_{s \to \infty}\, \langle \psi,U(s\cdot k)\psi'\rangle = \langle
\psi,\O\rangle\langle \O,\psi'\rangle\,.
\end{equation}
\end{Prop}
\noindent
{\it Sketch of Proof: } Let $W_R$ be the right wedge region and
$\cA(W_R)$
 the associated von Neumann algebra. Let $\D^{it}$,
$t \in \bR$, be the modular group corresponding to $\cA(W_R),\O$. Then
a theorem by Borchers \cite[Thm. II.9]{Bor} establishes the relation
$$ \D^{it}U(s\cdot e_+)\D^{-it} = U({\rm e}^{-2\p t}s \cdot e_+) $$
for all $t,s \in \bR$. Consequently, one can apply the argument of
Prop.\ I.1.3 in \cite{Dri} to gain relation (2.1). We point out that
the mass gap assumption enters in that argument.
\begin{Lemma}
Let $E := 1 - |\O \rangle \langle \O |$ be the projection orthogonal
to the vacuum vector, and let $P_{\pm} := P_0 \pm P_1$. Let $\psi,\psi'
\in {\rm dom}\,(P_0)$. Then there exist vectors $\psi_{\pm} \in \cH$
such that
$$ \langle \psi, E \psi' \rangle = \langle \psi_{\pm}, E P_{\pm} \psi'
\rangle\,. $$
\end{Lemma}
\begin{proof} From the mass-gap assumption we obtain
\begin{equation}
\frac{1}{|p_{\pm}|^2} \le \frac{|p_0|^2}{m^2}
\end{equation}
for all $p = (p_0,p_1) \in {\rm sp}(P) \backslash \{0\}$, where
$p_{\pm} := p_0 \pm p_1$. We claim that the vectors $\psi_{\pm} :=
(P_{\pm})^{-1}E\psi$ exist (in the sense of the functional
calculus). Indeed, denoting the spectral measure of $P$ by $F$, eqn.\
(2.2) implies
\begin{eqnarray*}
\lefteqn{||\,(P_{\pm})^{-1}E\psi\,||^2 = \int_{{\rm sp}(P)
    \backslash \{0\}}\frac{1}{|p_{\pm}|^2}
 \langle \psi,dF(p)\psi\rangle} \\
& \le & \int_{{\rm sp}(P) \backslash \{0\}}\frac{|p_0|^2}{m^2}\langle
\psi,dF(p)\psi \rangle \ 
 \le \  \frac{1}{m^2} ||\,P_0\psi\,||^2\,.
\end{eqnarray*}
Thus, by the functional calculus,
$$ \langle\psi,E\psi'\rangle = \langle E\psi,E\psi'\rangle = \langle
P_{\pm} (P_{\pm})^{-1} E\psi,E\psi'\rangle = \langle \psi_{\pm},E
P_{\pm}\psi' \rangle\,, $$
where we used that $E$ commutes with $P_{\pm}$.
\end{proof}
\begin{Prop}
 Let $f \in \Coin(\bR^2)$ with $f \ge 0$, $\int
f(x)\,d^2x = 1$, and define $f_{x,\l}(y):= \l^{-2}f(\l^{-1}(y-x))$ so
that $f_{x,\l}$ approaches for $\l \to 0$ the delta-distribution
concentrated at $x$. Then for each pair $A,B \in \Ain$, the limit
$$ \langle A\O,\Tmn[x]B\O\rangle := \lim_{\l \to 0}\, \langle
A\O,\Tmn(f_{x,\l}) B\O \rangle $$
exists and defines a quadratic form on $\Ain\O \times
\Ain\O$. Moreover,
\begin{itemize}
\item[{\rm (a)}] \quad $\bR^2 \owns x \mapsto \langle A\O,
 \Tmn[x] B\O\rangle$ is
  $C^{\infty}$,
\item[{\rm (b)}] \quad $\langle U(a)A\O,\Tmn[x]U(a)B\O\rangle = \langle
  A\O,\Tmn[x - a]\,B\O \rangle$\,, \quad
 $x,a \in \bR^2$,
\item[{\rm (c)}] \quad $(1 + P_0)^{-\ell}\Tmn[x](1 + P_0)^{-\ell} :=
  \lim_{\l \to 0}\, (1 + P_0)^{-\ell}\Tmn(f_{x,\l})(1+P_0)^{-\ell}$
  \\[2pt]
is a bounded operator on $\cH$,
\item[{\rm (d)}] \quad $\langle A\O,\lb \Tmn[x],B\rb\O\rangle = 0$ \quad
for all $A \in \Ain$, $B \in \Ain(\cO)$ and $x \in \cO^{\perp}$,
\item[{\rm (e)}] \quad $\partial^{\m}\langle A\O,\lb \Tmn[x],B\rb \O \rangle
  = 0$, \quad $A,B \in \Ain$,
\item[{\rm (f)}] \quad $\int \langle A\O,\lb
  \Tmn[x^1e_1],B\rb\O\rangle e_0^{\m}\,dx^1 = \langle A\O,P_{\n}B\O
  \rangle$, \quad $A,B \in \Ain$.
\end{itemize}
\end{Prop}
This proposition is a fairly direct consequence of assumption (xi),
see \cite[Thm.\ 12.4.8]{BW} (cf.\ also references cited there). 
 The commutator is defined as difference of
quadratic forms:
$$ \langle A\O,\lb \Tmn[x],B\rb \O \rangle := \langle
A\O,\Tmn[x]B\O\rangle - \langle B^*A\O,\Tmn[x]\O \rangle\,. $$
Observe that the integrand in (f) is supported on a finite interval
because of (d). It should also be noted that $\Tmn[x]$ will in
general not exist as an operator.
\begin{Lemma} 
 Let $W$ be a wedge region, $B \in \Ain$, and $j \in
  \bN$. Then for each $\varepsilon > 0$ there is some $B_{\varepsilon}
  \in \Ain(W)$ such that 
$$ ||\,(1 + P_0)^j(B - B_{\varepsilon})\O\,|| < \varepsilon\,. $$
\end{Lemma}
The proof can be given along similar lines as the proof of
 \cite[Prop.\ 14.3.2]{BW}; we may therefore skip the details.
 The cyclicity of $\O$ for the algebras $\Ain(W)$ enters here.
 In combination with (b)
and (c) of Prop.\ 2.3 one obtains as a simple corollary:

For each wedge region $W$, any $A,B \in \Ain$ and given $\varepsilon
> 0$ there is some $B_{\varepsilon} \in \Ain(W)$ so that 
\begin{equation}
 |\langle A\O,\Tmn[x](B-B_{\varepsilon})\O\rangle | < \varepsilon
\end{equation}
holds uniformly in $x \in \bR^2$.
\subsection{Main result}
In the present section we state and prove our main result about ANEC
in quantum field theory on two-dimensional Minkowski spacetime.
\begin{Thm} Let $\{\cH,\cA,U,\O,\Tmn\}$ be a quantum field theory with
  energy-momentum tensor on two-dimensional Minkowski spacetime
  fulfilling the assumptions (i)--(xi) of Section 2.1.

Let $k$ be any non-zero lightlike vector in $\bR^2$ and let $A,B \in
\Ain$, $a \in \bR^2$. Then it holds that
$$ \lim_{r_{\pm} \to \infty} \, \int_{-r_-}^{r_+} \langle
A\O,\Tmn[s\cdot k + a]\,B\O \rangle k^{\m}\, ds = \langle A\O,P_{\n}
B\O \rangle\,. $$
\end{Thm}
\begin{Cor} This implies the ANEC for all vector states induced
 by the dense, translation-invariant
 set of vectors $\{\psi =  A\O$: $A \in \Ain\}$, corresponding
 to energetically strongly damped,
  local excitations of the vacuum:
$$ \lim_{r_{\pm}\to \infty} \, \int_{-r_-}^{r_+} \langle \psi,\Tmn[s \cdot k +
a]\,\psi\rangle k^{\m}k^{\n}\, ds = \langle \psi,k^{\n}P_{\n}\psi
\rangle \ge 0 $$
since $k$ is lightlike and since the relativistic spectrum condition
holds.
\end{Cor}
\begin{proof}
The proof proceeds in three simple steps. For simplicity of notation, we
will give the proof only for the case $k = e_+$, the proof for $k =
e_-$ is obtained by analogous arguments. In view of translation
covariance, it suffices also to consider only the case $a=0$.
 \\[6pt]
{\it 1) } We will first show that for all $C \in \Ain$ 
\begin{eqnarray}
\lim_{s \to \pm\infty}\,\langle C\O,\Tmn[s\cdot e_+]\O \rangle& =&
0\,, \\
\lim_{r_{\pm}\to\infty}\, \int_{-r_-}^{r_+} \langle C\O,\Tmn[s \cdot e_+]\O
\rangle \,ds &=& 0\,.
\end{eqnarray}
To this end, let
\begin{eqnarray*}
\psi &:= & (1+P_0)^{\ell + 1}C\O\,,\\
\psi_{\m \n}' & : = & (1+ P_0)^{-(\ell + 1)}\Tmn[0](1 + P_0)^{-\ell}\O\,.
\end{eqnarray*}
One can see from Prop.\ 2.3(c) that $\psi,\psi_{\m\n}' \in {\rm
  dom}\,(1 + P_0)$. Moreover, denoting by $E := 1 -|\O\rangle \langle
\O |$ the projection orthogonal to the vacuum, we deduce, upon using
assumption (vii) (implying $\langle \O,\Tmn[x]\O\rangle = 0$) and
Prop.\ 2.3(b) together with the fact that $E$ commutes with $U(a)$, $a
\in \bR^2$, that
$$ \langle C\O,\Tmn[s\cdot e_+]\O \rangle = \langle \psi,E\,U(s\cdot
e_+)\psi_{\m\n}'\rangle \,, \quad s \in \bR\,.$$
Then relation (2.4) follows from weak asymptotic lightlike clustering,
Prop.\ 2.1. Furthermore,
by Lemma 2.2 it follows that there is a vector $\psi_+ \in \cH$ so
that
$$ \langle \psi,E\,U(s\cdot e_+)\psi_{\m\n}'\rangle = \langle \psi_+,E
P_+U(s\cdot e_+)\psi_{\m\n}' \rangle = \frac{1}{i}\frac{d}{ds}\langle
\psi_+,E\,U(s\cdot e_+)\psi_{\m\n}'\rangle \,.$$
\newpage\noindent
${}$ \par \vspace*{-1.5cm} \noindent
Thus one obtains
\begin{eqnarray*}
\lefteqn{\int_{-r_-}^{r_+}\langle C\O,\Tmn[s\cdot e_+]\O\rangle\,ds 
 = \int_{-r_-}^{r_+}
 \frac{1}{i}\frac{d}{ds}\langle\psi_+,E\,
U(s\cdot e_+)\psi_{\m\n}'\rangle\,ds}\\
& = &
\frac{1}{i}\left(\langle\psi_+,E\,U(r_+\cdot e_+)\psi_{\m\n}'\rangle -
\langle\psi_+,E\,U(r_-\cdot e_+)\psi_{\m\n}'\rangle
\right)
\end{eqnarray*}
and the last expression tends to 0 in the limit $r_{\pm} \to \infty$
in view of weak asymptotic lightlike clustering, Prop.\ 2.1. This
establishes relation (2.5).
\\[6pt]
{\it 2) } Relation (2.5) shows, for any $A,B \in \Ain$,
$$ \lim_{r_{\pm} \to \infty}\, \left(
 \int_{-r_-}^{r_+}\langle A\O,\Tmn[s\cdot e_+]B\O\rangle\,ds\  -
 \int_{-r_-}^{r_+} \langle A\O,\lb\Tmn[s\cdot e_+],B\rb\O\rangle\,ds
\right) = 0 $$
and hence, to prove the theorem, it suffices to demonstrate
\begin{equation}
\lim_{r_{\pm} \to \infty}\,\int_{-r_-}^{r_+}
\langle A \O,\lb \Tmn[s\cdot e_+],B \rb \O
\rangle e_+^{\m}\,ds = \langle A\O,P_{\n} B\O \rangle\,.
\end{equation}
To show this, we fix any $A,B \in \Ain$ and use the abbreviation
$$ \tmn(x) := \langle A\O,\lb \Tmn[x],B \rb \O \rangle\,. $$
Now we define two maps with values in $\bR^2$,
$$ h_+(s,\r) := s\cdot e_+ - \r\cdot e_-\,, \quad \ h_-(s,\r) := -s\cdot
e_+ + \r \cdot e_-\,, \quad s,\r \ge 0\,,$$
and the two triangle-shaped regions
\begin{eqnarray*}
 R_{+,r_+} &:=& \{h_+(s,\r): 0 \le s \le r_+,\ 0 \le \r \le s\}\,,\\
R_{-,r_-} &:= &\{h_-(s,\r): 0 \le s \le r_-,\ 0 \le \r \le
s\}\,.
\end{eqnarray*}
The region $R_{+,r_+}$ is bounded by the two lightlike line segments
$L_+(r_+) :=\{s\cdot e_+ : 0 \le s \le r_+\}$ and $H_+(r_+):=
\{h_+(r_+,\r): 0 \le \r \le r_+\}$, and by the spacelike line segment
$S_+(r_+) := \{x^1e_1 : 0 \le x^1 \le \sqrt{2}\,r_+\}$. Similarly,
$R_{-,r_-}$ is bounded by the line segments $L_-(r_-) := \{- s \cdot
e_+: 0 \le s \le r_-\}$, $H_-(r_-):= \{h_-(r_-,\r): 0 \le \r \le r_-\}$,
and 
$S_-(r_-):= \{-x^1e_1 : 0 \le x^1 \le \sqrt{2}\,r_-\}$. (Cf.\ Figure
1.) 
\begin{center}
\epsfig{file=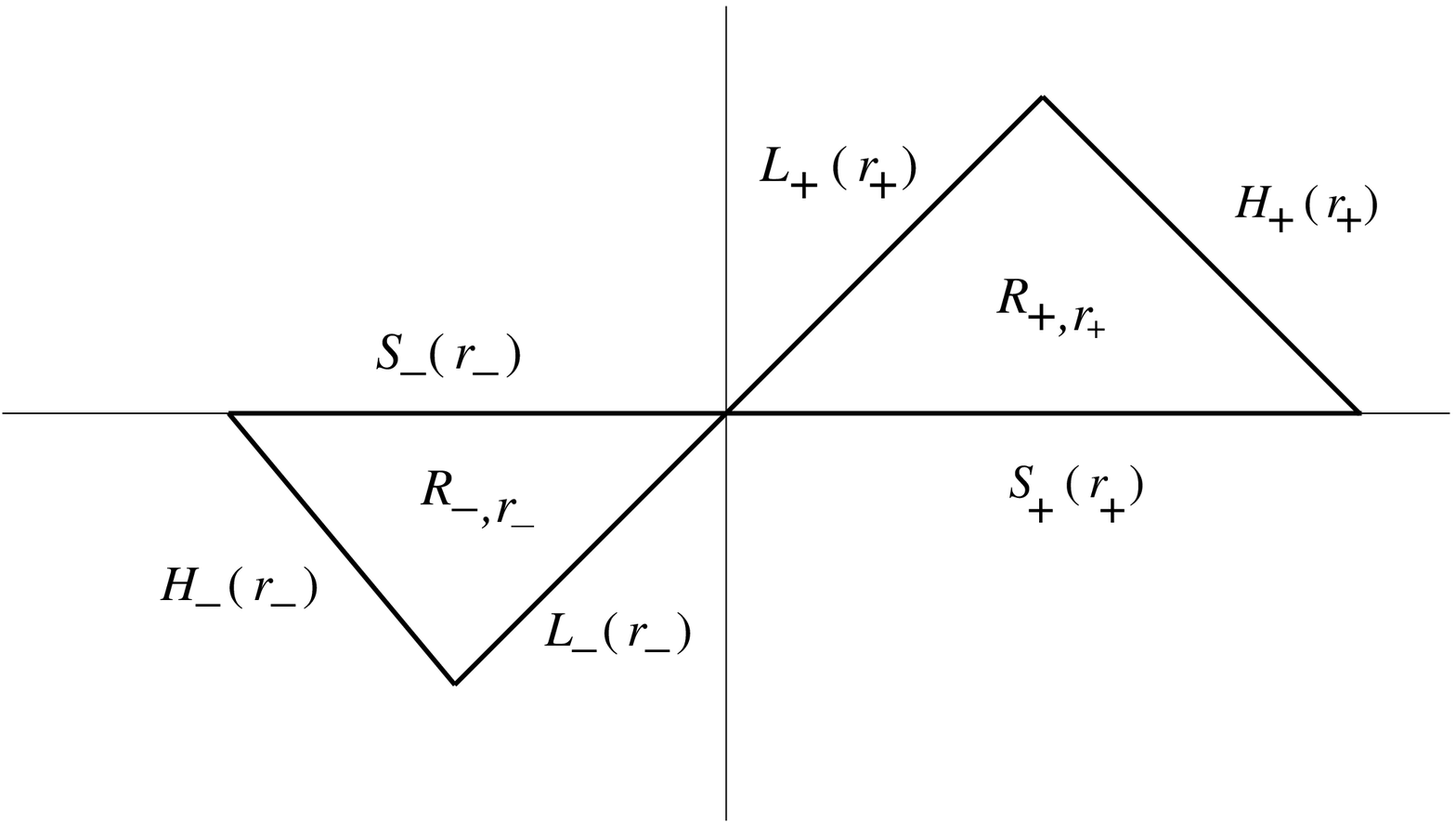, width=11.0cm}\\
{\small {\bf Figure 1. } \quad Sketch of the regions and  bounding
  line segments described in the text.}
\end{center}
Now we use $\partial^{\m}\tmn(x) = 0$ and thus, applying Gau{\ss}' law
 to the region $R_{+,r_+}$, we convert the integral of
$ v^{\m}=\t^{\m}{}_{\n}$ paired with the outer normal along
$L_+(r_+)$ into a sum of two integrals of $v^{\m}$ paired with
the inner normals along $H_+(r_+)$ and $S_+(r_+)$. Doing the same with
respect to the region $R_{-,r_-}$ (with the roles of inner and outer
normals interchanged) yields, with the above parametrizations of the
various line segments inserted,
\begin{eqnarray}
\lefteqn{ \int_{-r_-}^{r_+} \tmn(s\cdot e_+)e^{\m}_+\,ds =
  \int_{-\sqrt{2}\,r_-}^{\sqrt{2}\,r_+}\tmn(x^1e_1)e^{\m}_0\,dx^1} \\
& & - \int_0^{r_+}\tmn(h_+(r_+,\r))e_-^{\m}\,d\r\ 
 -\int_{0}^{r_-}\tmn(h_-(r_-,\r))e^{\m}_+ \,d\r \,. \nonumber
\end{eqnarray}
In view of Prop.\ 2.3(d,f), we deduce that the first integral on the
right hand side of (2.7) equals $\langle A\O,P_{\n}B\O\rangle$ as soon
as $r_+$ and $r_-$ are large enough. This implies that (2.6), and
hence the statement of the theorem, is proved once it is shown that
the two remaining integrals on the right hand side of (2.7) vanish in
the limit $r_{\pm} \to \infty$.
\\[6pt]
{\it 3) } The remaining step in the proof is therefore to show 
\begin{equation}
 \lim_{r_{\pm} \to \infty}\, \int_0^{r_{\pm}}\tmn(h_{\pm}(r_{\pm},\r))
\,d\r = 0\,.
\end{equation}
We will demonstrate this only for the ``$+$'' case, the reasoning for
the ``$-$'' case is similar.

It holds that $B \in \Ain(\cO_I)$ for $I =\{x^1e_1: |x^1| <
\sqrt{2}\,\xi\}$ with some sufficiently large $\xi > 0$. By Prop.\
2.3(d), $\tmn(x) = 0$ for $x \in (\cO_I)^{\perp}$, implying that
\begin{equation}
\int_0^{r_+}\tmn(h_+(r_+,\r))\,d\r = \int_0^{\xi}\tmn(h_+(r_+,\r))\,d\r
\,,
\end{equation}
i.e.\ the integral extends for all $r_+ > 0$ only over a fixed
interval of finite length.

Now choose some wedge region $W$ in the causal complement of $\bigcup_{r_+
  \ge 0}H_+(r_+) \subset W_R$, and let $\d > 0$ be arbitrary. According to
(2.3), one can find some $B_{\d} \in \Ain(W)$ so that 
$$ |\langle A\O,\Tmn[x](B - B_{\d})\O \rangle
| < \frac{\d}{2\xi} $$
uniformly in $x \in \bR^2$. Then
 $\langle A\O,\lb\Tmn[x],B_{\d}\rb\O\rangle = 0$ for all
$x \in H_+(r_+)$, and
\begin{eqnarray*}
\lefteqn{\int_0^{\xi}\tmn(h_+(r_+,\r))\,d\r = \int_0^{\xi} \langle
  A\O,\Tmn[h_+(r_+,\r)](B - B_{\d})\O\rangle\,d\r}\\
&+ & \int_0^{\xi}\langle(B^*_{\d} -
B^*)A\O,\Tmn[h_+(r_+,\r)]\O\rangle\,d\r\,.
\end{eqnarray*}
 The absolute value of the first integral on the right
hand side of the last equation can be estimated by $\xi\cdot\d/2\xi =
\d /2$. Owing to (2.4), the other integral on the right hand side of
the last equation converges to 0 for $r_+ \to \infty$ (note that the
integrands are bounded uniformly in $r_+$). Therefore we can find for
the given $\d > 0$ some $r > 0$ so that
$|\int_0^{\xi}\tmn(h_+(r_+,\r))\,d\r| < \d$ for all $r_+ > r$. By
(2.9), this establishes the required relation (2.8), and thus the
proof is complete.
\end{proof} 
\section{A result for lightlike half-lines}
\setcounter{equation}{0}
In this section we present a result indicating that ANEC fails to hold
in general for dense subsets of the vectors considered in Theorem 2.5
when the expectation value of the energy-momentum tensor is integrated
only over a lightlike half-line. This is of course no surprise in view
of the fact that a lightlike half-line has a large causal complement
together with the assumed properties of the energy-momentum
tensor. The precise formulation of the result is as follows.
\begin{Prop}
Let $\{\cH,\cA,U,\O,\Tmn\}$ be a quantum field theory with
energy-momen\-tum tensor on two-dimensional Minkowski-spacetime with the
properties assumed in Section 2.1. Let $k$ be a non-zero lightlike
vector in $\bR^2$, $a \in \bR^2$, and let $\cO$ 
be a double cone lying in the causal complement of the lightlike
half-line $L := \{s\cdot k + a : s \ge 0\}$.

Suppose that $\Ain(\cO)\O$ is dense in $\cH$ (Reeh-Schlieder property)
and that for all $A \in \Ain(\cO)$ there holds
$$ \liminf_{r \to \infty}\,\int_0^r \langle A\O,\Tmn[s\cdot k +
a]A\O\rangle k^{\m}k^{\n}\, ds \ge 0\,.$$
Then the Hilbertspace $\cH$ is one-dimensional and spanned by the
vacuum vector $\O$, and $\Tmn(f) = 0$ for all $f \in \Coin(\bR^2)$.
\end{Prop}
\begin{proof}
We consider only the case $k = e_+$ and $a = 0$, the general case is
proved analogously.

Then we observe that 
\begin{equation}
   \langle A\O,T[L]B\O \rangle := \lim_{r \to
     \infty}\,\int_0^r\langle A\O,\Tmn[s\cdot e_+]A\O\rangle
   e_+^{\m}e_+^{\n}\,ds = i\langle A\O,\Tmn[0]B\O\rangle
   e_+^{\m}e_+^{\n}
\end{equation}
holds for all $A,B \in \Ain(\cO)$ as can be seen from (2.5) together
with the fact that $\langle A\O,\lb\Tmn[s\cdot e_+],B\rb\O\rangle = 0$, $s
\ge 0$. Equation (3.1) defines a quadratic form $\langle
\,.\,,T[L]\,.\,\rangle$ on $\Ain(\cO)\O \times \Ain(\cO)\O$ which is by
assumption positive, i.e.\ $\langle A\O,T[L]A\O\rangle \ge 0$, $A \in
\Ain(\cO)$. It follows that there is an essentially selfadjoint,
positive operator $T_L^{1/2}$ with domain $\Ain(\cO)\O$ so that
$$ \langle T_L^{1/2}A\O,T^{1/2}_L B\O\rangle = \langle
A\O,T[L]B\O\rangle = \langle B^*A\O,T[L]\O\rangle\,, \quad A,B \in
\Ain(\cO)\,.$$
Using $\langle \O,T[L]\O\rangle = 0$ this implies $\langle
A\O,T[L]B\O\rangle = 0$ and hence, by (3.1),
\begin{eqnarray*}
\lefteqn{\langle(1+P_0)^{\ell}A\O,(1+P_0)^{-\ell}\Tmn[0]
(1+P_0)^{-\ell}(1+P_0)^{\ell}B\O\rangle e_+^{\m}e_+^{\n} }\\
 &=& \langle A\O,\Tmn[0]B\O\rangle e_+^{\m}e_+^{\n} \ = \ -i\langle
 A\O,T[L]B\O\rangle \ = \ 0
\end{eqnarray*}
for all $A,B \in \Ain(\cO)$. The set of vectors $(1+P_0)^{\ell}A\O$, $A \in
\Ain(\cO)$ is dense in $\cH$, therefore, using also covariance
(Prop. 2.3(b)), one arrives at
$$ (1+P_0)^{-\ell}\Tmn[x](1 + P_0)^{-\ell}e_+^{\m}e_+^{\n} = 0\,,
\quad x \in \bR^2\,.$$
Thus $\langle A\O,\Tmn[x]B\O\rangle e_+^{\m}e_+^{\n} = 0$ for all $A,B
\in \Ain$ and $x \in \bR^2$, and in view of Theorem 2.5, this entails
$$ (P_0 + P_1)B\O = 0\,, \quad B \in \Ain\,.$$
Since we have imposed the mass gap assumption (iv), we may apply
Proposition I.1.2 of \cite{Dri} to conclude that this is only possible
if $B\O$ is parallel to the vacuum vector $\O$. As the set of vectors
$\Ain\O$ is dense in $\cH$, this implies $\cH = \bC\cdot \O$, and by
the vanishing of the vacuum-expectation value of $\Tmn$, finally
$\Tmn(f) = 0$ for all $f \in \Coin(\bR^2)$.
\end{proof}
\section{Concluding remarks}
It has been shown that in two-dimensional Minkowski spacetime the ANEC
can be derived under very general hypotheses for quantum field
theories endowed with an energy-momentum tensor. The
two-dimensionality was quite essential in exploiting the vanishing of
the divergence of $\Tmn$ in the proof of Thm.\ 2.5 and it is not at
all clear if our simple argument can be generalized to higher
dimensions. So the general validity of ANEC in higher dimensions
remains unsettled.

Concerning quantum field theory in curved spacetime, a familiar
problem is that there are no candidates for a vacuum state of a
quantum field theory owing to the circumstance that in general there
are no spacetime symmetries. Then already the characterization of
physical states and the definition of expectation values of an
energy-momentum tensor poses considerable problems (see \cite{WaldII}
for discussion how this problem is treated in the case of free
fields). Clearly the the question if, and in which sense, ANEC may
hold in quantum field theory in curved spacetime is connected to this
circle of problems, particularly to the issue of how to characterize
states which may be viewed as playing the role of preferred,
vacuum-like states.  Here ANEC is in some ways attractive as imposing
a global constraint on candidates for such states, complementary
to other prominent conditions, like the Hadamard condition
(cf.\ \cite{WaldII}) or the microlocal spectrum condition (see
\cite{BFK}) which constrain the short-distance properties of physical
states. A drawback is that ANEC is a condition which cannot be tested
locally. It is to be hoped that more progress in understanding the
relation between the said local conditions and ANEC will be made in
the future. 
\\[18pt]
{\bf Acknowledgement.} It would like to thank D.\ Buchholz for useful
comments on the topic.

\end{document}